# Experimental investigation of the local environment and lattice distortion in refractory medium entropy alloys

Lola Lilensten[1,2*], Karine Provost[2], Loïc Perrière[2], Emiliano Fonda[3], Jean-Philippe Couzinié[2], Fabienne Amann[1], Martin Radtke[4], Guy Dirras[5], Ivan Guillot[2]

[1] Chimie ParisTech, PSL University, CNRS, Institut de Recherche de Chimie Paris, 75005 Paris, France

[2] Université Paris Est Créteil, CNRS, ICMPE, UMR7182, Thiais F-94320, France

[3] Synchrotron SOLEIL, L'Orme des Merisiers, Saint Aubin BP 48, F-91192 Gif S Yvette, France

[4] BAM Bundesanstalt für Materialforschung und -prüfung, Richard-Willstätter Strasse 11, D-12489 Berlin, Germany

[5] Université Paris 13, Sorbonne Paris Cité, CNRS, LSPM (UPR 3407) 99 avenue JB Clément, 94430 Villetaneuse, France

* corresponding author : lola.lilensten@cnrs.fr

**Abstract**

EXAFS analysis of pure elements, binary and ternary equiatomic refractory alloys within the Nb-Zr-Ti-Hf-Ta system is performed at the Nb and Zr K-edges to analyze the evolution of the chemical local environment and the lattice distortion. A good mixing of the elements is found at the atomic scale. For some compounds, a distribution of distances between the central atom and its neighbors suggests a distortion of the structure. Finally, analysis of the Debye-Waller parameters show some correlation with the lattice distortion parameter $\delta^2$, and allows to quantify experimentally the static disorder in medium entropy alloys.

**Manuscript**

Numerous studies have now evidenced that multicomponent concentrated alloys with a single solid solution phase, named "high entropy alloys" (HEAs) or "complex concentrated alloys" (CCAs), can be obtained. Due to the concept itself of mixing several elements with different radii, the local atomic structure is of utmost interest [1]. Indeed, a supposed severe lattice distortion has been extensively used to explain the properties of these materials, but investigation of such local effects is challenging.

*Ab-initio* methods have been used, showing in most cases that long-range interactions are found. More precisely, body-centered cubic (bcc) refractory compositions, such as NbTaTiV, NbMoTaW, HfNbZr, HfNbTiZr, HfNbTaTiZr show a rather strong lattice distortion [2–4]. This effect seems more pronounced in refractory bcc compositions than in face-centered cubic (fcc) 3d CCAs [2]. Surprisingly, so-called "medium entropy alloys" (MEAs) such as HfNbZr are found to display a more substantial distortion compared to the five elements alloy HfNbTaTiZr.

Experimental validation of the lattice distortion suggested by calculation requires advanced techniques. Advanced high-resolution transmission electron microscopy (HRTEM) has recently been used to prove the existence of short-range ordering (SRO) in a MEA CrCoNi [5]. This had been suggested first by X-ray adsorption measurements, evidencing that Cr bonds more favorably with Ni and Co [6]. These results are even more valuable as no structural distortion was evidenced by X-ray and neutron total scattering experiments [6]. Similar X-ray absorption studies have been performed on $Al_8Cr_{17}Co_{17}Cu_8Fe_{17}Ni_{33}$ fcc alloy, revealing the presence of SRO and a variation of the electronic structure [7]. Comparing the Debye-Waller coefficient obtained by these two techniques on the CrMnFeCoNi composition measured at 50K and at room temperature allowed for the first estimation of the intrinsic static disorder due to the atomic size mismatch [8]. A similar approach using neutron total scattering measurements was performed by Owen et al. [9] in Ni-based alloys with increasing complexity: pure Ni was compared to Ni-20Cr, Ni-25Cr, Ni-22Cr, Ni-37.5Co-35Cr, and the equiatomic CrMnFeCoNi alloy, revealing a variation in the Debye-Waller coefficients. However, the authors rationalize their findings by a mere variation in the homologous temperatures.

Facing this experimental knowledge on fcc CCAs, and based on the *ab-initio* suggestion that bcc HEA would present a stronger lattice distortion [2], this latter family lacks systematic experimental investigations. Among the few studies that looked at them, Maiti et al. measured a high degree of lattice distortion, as well as Zr short-range clustering after annealing in the ZrNbHfTa system [10]. Lattice distortion was also observed by Zou et al. in NbMoTaW by HRTEM [11]. Finally, a pair distribution function (PDF) study by XRD and neutron diffraction on a MEA ZrNbHf also showed lattice distortion, as atoms' local environment are described with 15.5 nearest neighbors (NN) instead of the 14 first and second neighbors of the bcc structure [12].

This study thus proposes to systematically investigate by EXAFS the evolution of the local environment in MEAs, that are compared to the pure elements. Alloys with higher complexity (4- or 5-elements) are not considered here since the number of parameters to fit would be too large for the available signal. Besides, ternary alloys may display a stronger lattice distortion than 5-elements alloys [2]. The compositional system Ti, Zr, Nb, Hf, Ta was chosen, as the 5-elements alloy based on these elements benefits from several studies suggesting that it forms a ductile solid solution [13–17].

Zr and Nb were chosen as references and represent two distinct crystalline structures at room temperature, namely hexagonal closed packed (hcp) and bcc. Besides, the choice of these two elements as references was guided by the accessibility of their K-edge energy, and the fact that there is no overlap between their K-edge and other edges, so that the signals can be analyzed without artifact.

Binary alloys and ternary alloys listed in table 1 were considered. All alloys have equiatomic compositions, which allows to characterize the local environment of Zr and of Nb in alloys of increasing chemical complexity.

*Table 1: alloys investigated. In bold character are the compositions that were measured both at the Zr K-edge and at the Nb K-edge. The alloys with hcp structure are italicized, the other ones have a bcc structure.*

| Edge | Pure metal | Binary alloy | Ternary alloy |
| --- | --- | --- | --- |

| Zr K-edge (17.998 keV) | Zr | *ZrTi* *ZrHf* | *ZrTiHf* **NbTiZr** |
|---|---|---|---|
| Nb K-edge (18.986 keV) | Nb | NbTi NbTa NbHf | NbTiTa **NbTiZr** |

Ingots of 15g were prepared for the considered compositions. Arc-melting was performed in an Ar-atmosphere on a water-cooled copper plate. The chamber was flushed twice with Ar prior to melting, and a Ti-Zr getter was used before melting the alloys. Master ingots were prepared for alloys with three arc-melted elements separately three times and then arc-melted together (2 melting steps), the specimens were flipped between these two steps. Alloys with only two elements were directly melted in 3 fusions steps. Induction melting was then performed for all the alloys in a water-cooled sectorized copper mold under helium to ensure chemical homogeneity in the ingots. Finally, the ingots were cast by arc-melting to get a cylindrical shape. Alloys then underwent thermomechanical treatments described in the supplemental material (table S1). XRD was finally performed with a Panalytical X'Pert Pro diffractometer using the Co $K_\alpha$ radiation, to confirm that the structure was the expected one (hcp or bcc) and that the alloys were a single phase. Lattice parameters were extracted by Rietveld refinement using Maud [18]. Their composition was also confirmed by EPMA. Details are provided in table 2, and additional Rietveld refinement parameters can be found in supplementary materials (Table S2, and Figure S1). Specimens of optimal thicknesses regarding the absorption were prepared by manual polishing on SiC papers with grades 320 to 4000 (exact thickness detailed in supplementary material, table S1). The thin foils were then taped in between two bands of Kapton tape. A minimum of two samples per composition were produced.

Synchrotron measurements were carried out at the BAMline at Helmholtz-Zentrum Berlin in transmission mode at Zr and Nb K-edges. The recording was done with a Si 111 monochromator, using a 1eV step for the XANES. 3 to 6 spectra were acquired for each sample to minimize statistical error (statistical error and error bars evaluated as recommended by the IXS Standard and Criteria Subcommittee [19]). The XANES normalization and EXAFS extraction were done using MAX-Cherokee [20], and EXAFS fits were done with MAX-Round Midnight [20], using a 3-12

Å$^{-1}$ range and a k$^3$ weighting for all samples (more details below). For sample NbZrTi we used the ifeffit code [21] to simultaneously simulate Nb and Zr edges to naturally restrain parameters like Zr-Nb distance and Debye Waller terms. We used theoretical phases and amplitudes calculated by FEFF8 code [22].

*Table 2: composition probed by EPMA (at.%), experimentally measured lattice parameters (XRD), and published lattice parameters of the investigated alloys (from [23–31]).*

| Sample | Hf | Nb | Ta | Ti | Zr | Exp. lattice parameters (Å) | Reported lattice parameters (Å) |
|---|---|---|---|---|---|---|---|
| Nb | - | 100 | - | - | - | $a$ = 3.307 | $a$ = 3.300 |
| Zr | - | - | - | - | 100 | $a$ = 3.239<br>$c$ = 5.135 | $a$ = 3.232<br>$c$ = 5.147 |
| NbTi | - | 49.9 ± 2.4 | - | 51.1 ± 2.4 | - | $a$ = 3.287 | $a$ = 3.286 |
| NbHf | 50.1 ± 0.5 | 49.9 ± 0.5 | - | - | - | $a$ = 3.423 | $a$ = 3.4121 |
| NbTa | - | 50.2 ± 3.2 | 49.8 ± 3.2 | - | - | $a$ = 3.309 | $a$ = 3.300 |
| ZrTi | - | - | - | 50.6 ± 0.1 | 49.4 ± 0.1 | $a$ = 3.116<br>$c$ = 4.907 | $a$ = 3.12<br>$c$ = 4.90 |
| ZrHf | 51.0 ± 0.3 | - | - | - | 51.0 ± 0.3 | $a$ = 3.223<br>$c$ = 5.109 | $a$ = 3.209<br>$c$ = 5.126 |
| NbTiTa | - | 33.3 ± 0.2 | 32.7 ± 0.9 | 34.0 ± 0.8 | - | $a$ = 3.298 | $a$ = 3.285 |
| ZrTiHf | 34.2 ± 0.9 | - | - | 34.6 ± 1.1 | 31.2 ± 1.8 | $a$ = 3.139<br>$c$ = 4.971 | / |

| NbTiZr | - | 31.9 ± 2.6 | - | 34.4 ± 0.7 | 33.7 ± 1.9 | $a = 3.389$ | $a = 3.401$ |

Figure 1 shows the XANES spectra of all the compounds at the Zr K-edge (Figure 1a) and Nb K-edge (Figure 1c). The edge being at the same energy for all the compounds (no shift observed), thus it is concluded that no change happens in the Zr and the Nb electronic structure and oxidation state. Therefore, the correction of the edge energy ΔE was determined for the pure compounds and applied to all the alloys measured at the corresponding edges. The EXAFS spectra are given in Figure 1b and Figure 1d for the Zr K-edge and the Nb K-edge, respectively. For both edges, similar results are observed: although clear oscillations are maintained up to 12 Å$^{-1}$, the signal is attenuated for some binary and ternary compounds. This may suggest that for some compounds such as ZrHf or NbHf, the presence of heavy diffusers might lead to destructive interferences between the signals or a higher disorder.

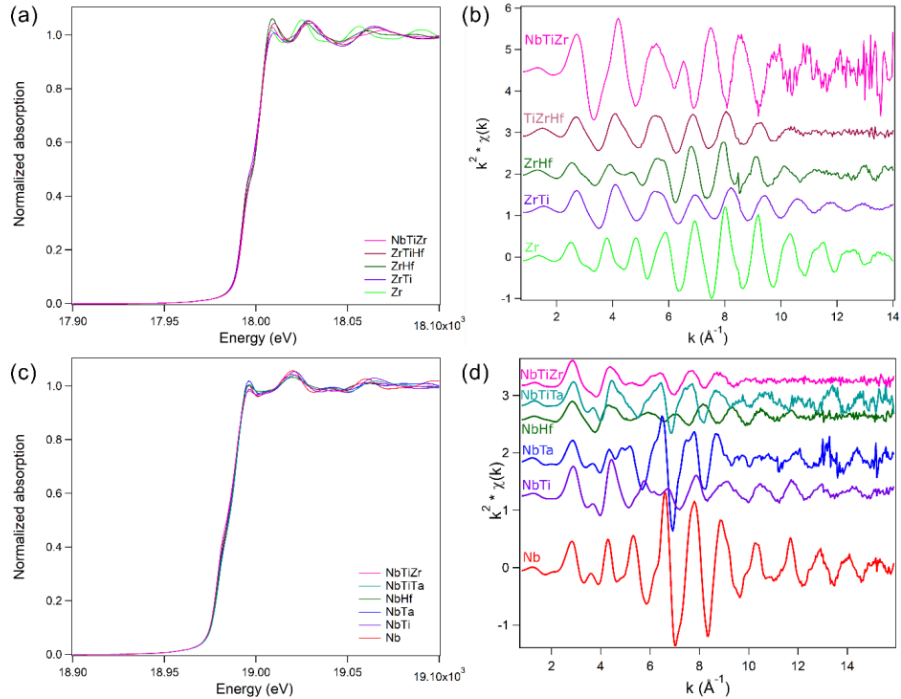

*Figure 1: (a) XANES spectra at the Zr K-edge zoomed in around the edge. (b) Extracted EXAFS spectra at the Zr K-edge. (c) XANES spectra at the Nb K-edge zoomed in around the edge (d) Extracted EXAFS spectra at the Nb K-edge.*

Considering the available range of signal, the first coordination sphere was considered in the following with the analysis of the 12 NN in alloys with hcp alloys (at a distance of *a*) and the 8+6 NN for bcc alloys (distance of $\frac{a\sqrt{3}}{2}$ and *a*, respectively), identified by the subscripts 1 (1st NN) and 2 (2nd NN). Although in the hcp structure, the 12 neighbors are not strictly at equal distances due to the *c/a* ratio, the difference in distances is too small to be picked up by the EXAFS spatial resolution (0.13 Å for a signal up to $k_{max} = 12$ Å$^{-1}$). However, the resolution is good enough to differentiate between the 1st NN and the 2nd NN of the bcc structure. The number of each neighbor, their distance (considering one shell for the hcp compounds and two shells for the bcc ones), and the Debye-Waller (DW) parameter σ² were fitted for all the studied alloys. When needed, some parameters remained constrained to ensure convergence of the fit, such as the total number of NN in the first or the second shell (see legend of Figure 2b and details in supplemental materials). Examples of EXAFS fits are plotted in supplementary materials (Figure S2).

The results of the best fits for each compound are presented in Figure 2 and Figure 3, the values presented in these Figures are also listed in supplemental materials together with details on possible constraints used during the fit procedure. Figure 2a shows the distance obtained between the central atom (Zr or Nb) and its neighbors (itself, or the alloying element(s)). Note that the error bars correspond to the fit's ones and they are not related to the above mentioned resolution of 0.13 Å that estimate the resolving power for identical scattering neighbors (e.g. two distinct Zr-Zr distance values instead of their weighted average associated with a larger Debye Waller factor). The distances between neighbors are systematically compared to the theoretical ones, calculated based on the atomic radii and indicated by crosses in the plot [32]. The Zr-Zr and Nb-Nb distances are in green and red, respectively, and the distance between the central atom X (X being Zr or Nb) and other elements are colored; X-Ti, X-Hf, X-Ta are in purple, dark green and blue, respectively.

The results of Figure 2a show that reasonable and similar distances are obtained between the central atom and its NN for most compounds (such as NbTa), yet, for some others, differences in distances depending on the nature of the neighbor are found, such as the NbHf compound, or ZrTiNb. Significant variations of NN distances depending on the NN's nature suggest that some alloys have a strong lattice distortion.

Figure 2b provides information on the chemical environment of the atoms. EXAFS fits of the pure metals show that the right amount of NN is found. Almost perfect mixing is obtained for the ZrTi, ZrHf, and ZrTiHf hcp compounds. Regarding the bcc compositions, mixing is always observed on the two shells. However, the mixing is found to be non-equiatomic (in NbHf and NbTa, Hf and Ta are slightly over-represented, when Ti is a bit under-represented in NbTi). This tendency is confirmed in NbTiTa. However, Ti seems well mixed in the case of NbTiZr. These results are unexpected, in the light of the mixing enthalpy values that are either equal to zero (Nb-Ta, Zr-Ti, Zr-Hf) or slightly positive (2 kJ.mol$^{-1}$ for Nb-Ti, 4 kJ.mol$^{-1}$ for Nb-Zr and Nb-Hf) [33]. Although not strictly reaching 1/2-1/2 or 1/3-1/3-1/3 mixing in all the cases, these results confirm that the binary and ternary alloys presented in this study display mixing of the elements at the atomic scale. Yet, the non-equiatomicity of the mixing and possible short range ordering, leading to very local compositional changes might be helpful to interpret some properties of the alloys, such as solid solution strengthening [34].

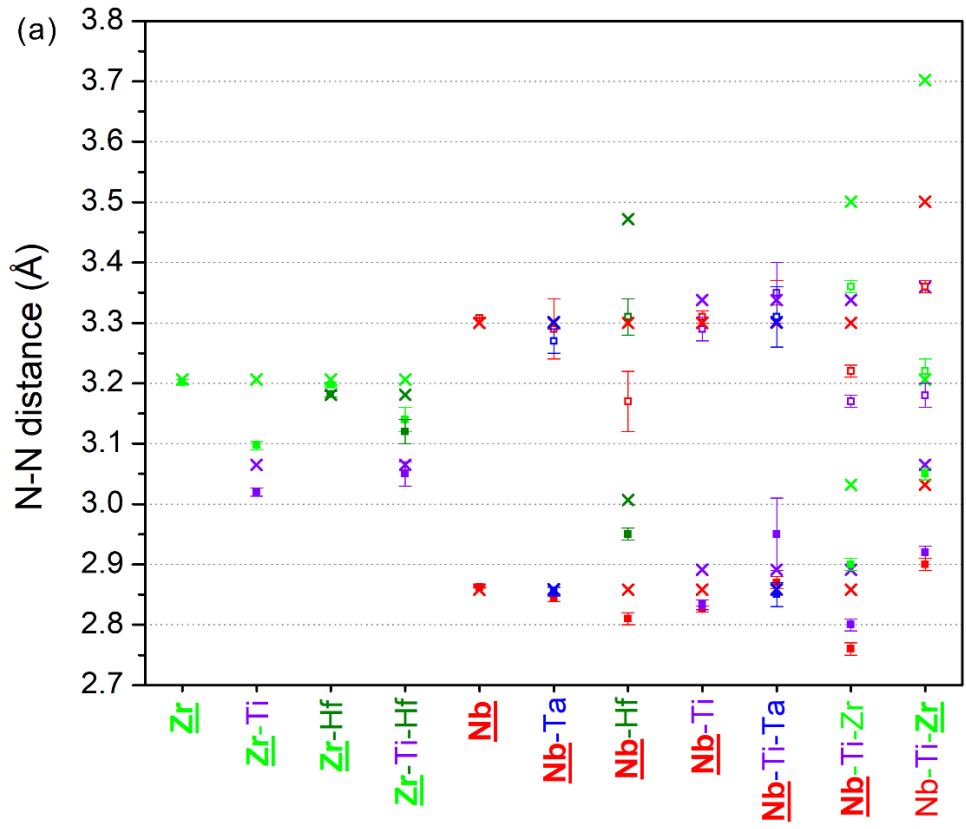

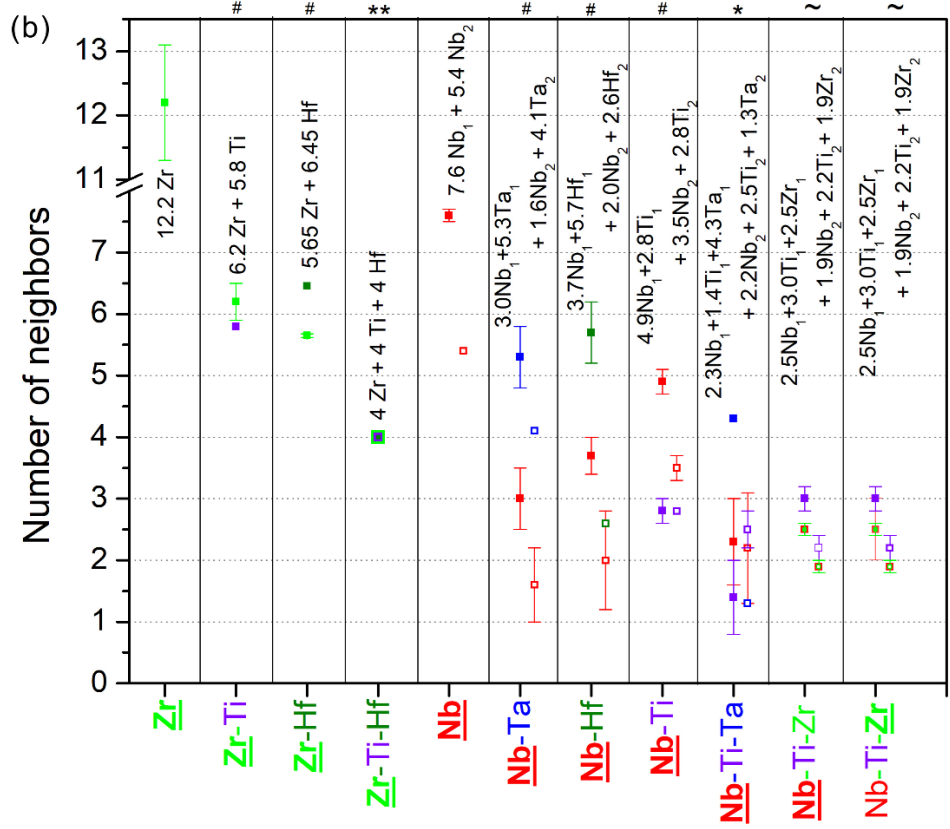

*Figure 2: (a) distance between the investigated atom (Zr or Nb, indicated by a bold and underlined character in the legend) and its neighbors. Crosses correspond to the theoretical distance based on a mixture law considering the sum of the atomic radii. For bcc structures, 1$^{st}$ NN are in plain squares and 2$^{nd}$ NN in open squares. The error provided is the one of the experimental fit. (b) Chemical composition around the investigated atom (in bold and underlined). For bcc compositions, the first NN are shown by plain squares, and the second NN by open squares. Some fitting constraints have been used: # indicates cases where the total number of neighbors sums up to 12 (hcp) or 14 (bcc); * where the first nearest neighbors sum up to 8 and the second nearest neighbors to 6; ** had imposed a number of 12neighbors and is the best fit (according to the QF) of a series of fits where the NN numbers varied; ~ is the result of the joined fit of ZrNbTi at both Nb- and Zr-K-edges simultaneously (details provided in supplementary materials)*

The DW parameter $\sigma^2$, that measures the dynamic (thermal) and static (structural) pair distribution widths commonly used as a measure of structural disorder, is considered next.

First, variation of $\sigma^2$ with respect to the $T/T_m$ ratio (T being the temperature of the experiment and $T_m$ the melting temperature, calculated with the TCHEA5 Calphad database) is plotted in Figure 3a. Indeed, the dynamic disorder, that is part of the $\sigma^2$ parameter, is related to the temperature. Although the melting temperatures of the alloys vary between 1837K (for ZrTi) and 3003 K (for NbTa), their $T/T_m$ ratio is only within the range 0.10 and 0.16. Figure 3a shows that NbTa and Nb, that have the highest melting temperatures and therefore the lowest $T/T_m$ ratio also logically have rather low $\sigma^2$ values, hence low dynamic or thermal disorder. However, as $T/T_m$ ratio increases (lower $T_m$), the $\sigma^2$ values taken by the various alloys do not display a clear trend and rather form a dispersed cloud of data. Therefore, it is hypothesized that although temperature can have an impact, the $T/T_m$ ratio range of all these refractory compounds is narrow enough to assume a similar thermal disorder in all the compounds. One can also notice that $\sigma^2$ takes values rather high (between 0.008 and 0.016 Å²) compared to values published for fcc compounds, which can be explained either by a tendency to overestimate the DW parameter in EXAFS compared to total scattering techniques, as observed for CrMnFeCoNi [8], or by a larger disorder in these refractory compounds, compared to 3d-elements alloys [2].

Next, the static disorder is addressed. In the field of HEAs, the atomic size mismatch parameter δ is conventionally used, and defined as $\delta^2 = 10^4 \Sigma_i c_i \left(1 - \frac{r_i}{\bar{r}}\right)^2$ (equation 2) [35], $r_i$ being the atomic radius, $c_i$ are the concentrations, $\bar{r} = \Sigma c_i r_i$ (equation 3) is the weighted average of atomic radii. These parameters are close by definition to EXAFS observables, where the pseudo DW factor is defined as $\sigma_i^2 = \langle (r_i - R_i)^2 \rangle$ (equation 4) where $r_j$ is the instantaneous value of the half scattering path (i.e. distance) $R_j$. An average DW over all similar scattering paths can then be defined as $\sigma^2 = \Sigma_i \Sigma_j c_i c_j \sigma_{i,j}^2$ (Equation 5), where DW is averaged over all pair distributions depending on the concentration of the absorber and the scatterer. This value approximates what obtained in this work since most atomic pairs have been probed and an average DW has been used. Finally, if differences of atomic radii translate into a globally broader pair distribution, we should expect that σ² scales as δ². σ² was thus plotted as a function of δ² in Figure 3b. The $r_i$ values used here are tabulated and $c_i$ was taken as the nominal composition of the element $i$ [35,36].

The plot shows indeed a relationship between the two parameters, although care must be taken considering the various hypotheses detailed above as well as the number of available points. Yet, the σ² coefficient linearly increases as a function of δ². Pure compounds and compounds with similar atomic radii for their constituting elements have low σ², whereas compounds such as TiZrHf or NbHf have DW coefficients up to twice as large. One can notice that this increase of the disorder with δ² is observed independently of the crystal structure, as bcc (close symbols) and hcp alloys (open symbols) follow the same trend and are aligned on the same line. Measurement of σ² thus allows to quantify a lattice distortion with a reliability stronger than that based on the variations of distances between nearest neighbors. However, it is interesting to notice that the alloys with largest DW coefficients, i.e. ZrTi, ZrTiHf, NbHf, and NbTiZr, are the ones that also show a meaningful difference between the NN distances measured by EXAFS and the theoretical ones, as well as a large dispersion between the distances, based on the nature of the NN (see Figure 2a), confirming the first hypothesis of a strong lattice distortion made with the results of Figure 2a.

This result of Figure 3b confirms that the atomic size mismatch increases the disorder at the local scale. It is also very interesting to notice that the disorder does not increase with the number of compounds, since binary alloys (square symbols) can have larger DW parameters than ternary alloys (triangles in Figure 3), see for instance TiNbTa and TiZr.

Finally, the DW parameter of the pure compounds, where the static disorder is supposed to be minimal, being chemically pure, provides an estimate of the thermal disorder baseline, that being hypothesized constant based on Figure 3a (evidenced as a grey region in Figure 3b). The increase of the σ² value represents the increased static disorder, meaning the alloy's distortion, as the static and dynamic contributions add up in σ². Therefore, the difference between the σ² parameter of an alloy and that of the pure metals allows estimating the distortion taking place in the alloy. Considering the error bars, and assuming a constant thermal disorder, it is suggested that an increase of up to 0.006 Å² can be obtained for the present set of alloys.

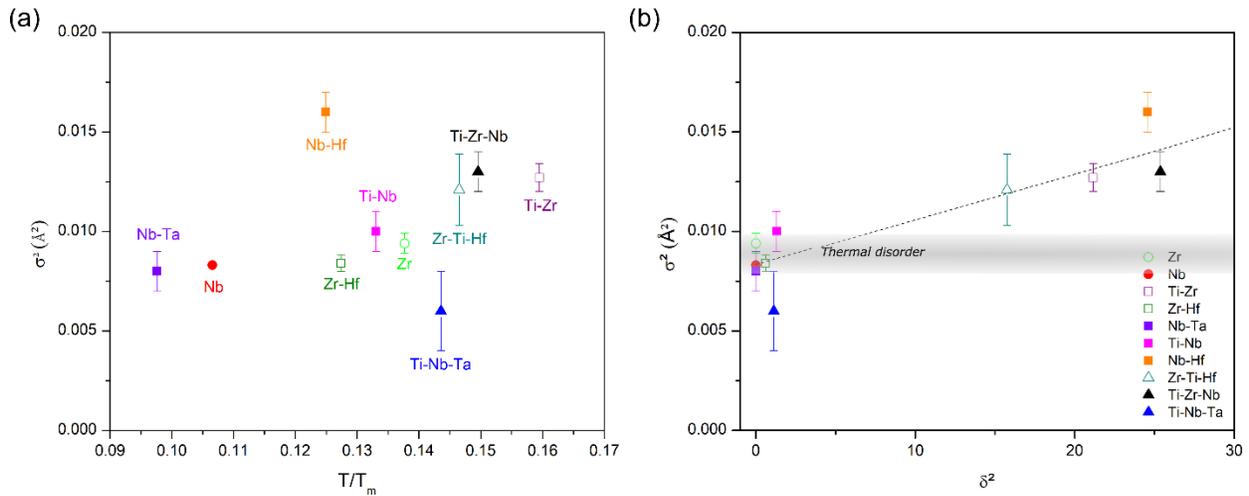

*Figure 3: (a) plot of the Debye-Waller parameter σ² as a function of $T/T_m$ (b) plot of the Debye-Waller parameter σ² as a function of δ, values calculated in the EXAFS fits. The grey area represents the thermal disorder. The dashed black line is a linear fit of the experimental points. Pure compounds are represented with circles, binary alloys with squares and ternary alloys with triangles. Open symbols correspond to compounds with hcp structure and plain symbols to compounds with bcc structure.*

In conclusion, by investigating the environment of Zr and Nb of 8 alloys and the pure metals by EXAFS, this study evidenced that mixing of the elements, sometimes non-equiatomic, is obtained down to the atomic scale in MEAs. Lattice distortion can be first approached roughly by comparing the X-X and X-Y distances, X being the central atom and Y the other element. When the X-X and X-Y distances between neighbors are strongly different, lattice distortion is expected, such as NbHf and NbTiZr. At last, the analysis of the DW parameter's evolution allows a physical measurement of the lattice distortion in refractory MEAs, which appears to be proportional to the theoretical parameter δ squared ($δ^2$), independently of the crystal structure. Notably these values are not increasing with the number of components, and they are not necessarily maximal for a largest number of alloying elements, going against the lattice distortion principle of HEAs. This last quantification with the DW parameter should be of great relevance to reach a better understanding of the HEAs properties and modeling purposes.

**Acknowledgements**:

We thank HZB for the allocation of synchrotron radiation beamtime.